\documentclass[a4paper,12pt]{article} 
\usepackage{graphicx}
\usepackage{times}
\usepackage{graphicx}
\usepackage{latexsym}
\usepackage{amsmath}
\usepackage{amssymb}
\usepackage{authblk}
\usepackage{setspace}
\begin{document}
\title{On the realizability of relativistic acoustic
geometry under a generalized perturbation scheme
for axisymmetric matter flow onto black holes}
\author[a]{Deepika B Ananda}
\author[b]{Sourav Bhattacharya}
\author[b]{Tapas K Das}
\affil[a]{\small Indian Institute of Science Education and Research, Pune, 411008, India.}
\affil[b]{\small Harish Chandra Research Institute, Chattnag Road, Jhunsi, Allahabad, 211019, India.\footnote{Current affiliation}}
\date{}
\maketitle
\begin{abstract}
\noindent
We propose a novel linear perturbation scheme to study the stability 
properties of the stationary transonic integral solutions for axisymmetric matter 
flow around astrophysical black holes for the Schwarzschild as well as for 
rotating Rindler spacetime. We discuss the emergence of 
the relativistic acoustic geometry as a consequence of such stability analysis. 
Our work thus makes a crucial connection between two apparently non-overlapping 
fields of research - the accretion astrophysics and the analogue gravity phenomena. 
\end{abstract}
\section{Introduction} 
\label{intro}
The stationary transonic accretion solutions have long been found 
useful to probe the spectral signature of astrophysical black holes, see, e.g., 
\cite{Frank King Book} for a detail review. It is, however, important to 
ensure the stability of such stationary solutions - at least within a reasonable 
astrophysical time scale - since transient phenomena are not quite uncommon 
during accretion processes. One can accomplish such task by perturbing the 
corresponding spacetime dependent fluid dynamic equations governing the 
accretion process and by studying whether such perturbation converges to 
ensure the stability of the transonic solutions of the time 
independent part of the aforementioned fluid dynamic equations. In the present work, 
we provide a linear perturbation scheme for low angular momentum 
irrotational inviscid accretion to ensure that the corresponding stationary 
integral transonic solutions are stable and to demonstrate that the relativistic 
acoustic geometry emerges from such perturbation analysis. Such accretion phenomena are observed in OB stellar winds accretion onto detached binary systems, semi-detached low-mass non-magnetic binaries and matter flow onto the supermassive black hole situated at the dynamical heart of our Galaxy. The linear stability analysis employed here helps us to determine whether the acoustic geometric structure (sonic or critical points etc.) present in almost any astrophysical accretion phenomenon, can in principle be observed. We also note that the stability of the accretion process or those stationary solutions may also be relevant in the context of the stability of the spacetime itself. Because if an accretion process `grows' unboundedly with time, its backreaction may change the background spacetime structure.

As of now, the acoustic geometry for any classical fluid configuration has been 
obtained by perturbing the corresponding velocity potential, see, 
e.g., \cite{Analogue Gravity Living Review Article} for 
further detail. We, for the first time in literature, obtained the
acoustic geometry for general relativistic axisymmetric accretion flow by perturbing
a more physically realizable, astrophysically relevant, and most importantly, observationally measurable
entity - the mass accretion rate.
Since the density and the velocity fields are directly interlinked through the mass accretion rate, perturbation of the 
accretion rate would provide a more physically realizable stability analysis 
scheme. Accretion rate is associated with matter flow onto astrophysical black holes 
and hence the aforementioned formalism shows that accreting black hole 
systems can be considered as an example of classical analogue system 
naturally found in the Universe and is unique in the sense that only in such 
systems both the gravitational as well as acoustic horizons exist. This is 
a very important finding as we believe since it shows how the actual gravitational field 
determines the salient features of the emergent gravity phenomena.  
 
In subsequent sections, we shall demonstrate the application of our perturbation scheme 
first for the Schwarzschild spacetime and then for rotating Rindler spacetime, 
describing the background fluid flow. 
\section{Axisymmetric accretion in the Schwarzschild metric} 
\label{schwarzschild}
We consider the Schwarzschild line element of the form 
$ds^{2}= -f dt^{2}+f^{-1}dr^{2}+r^{2}d  \theta^{2}+r^{2}{\rm sin}^{2}\theta d\phi^{2}$,
where $ f\equiv f(r) = 1-\dfrac{2}{r}$. For a dissipationless perfect fluid, the 
energy momentum tensor can be written as 
$T^{\mu \nu} = \left( \epsilon + p \right) v^{\mu} v^{\nu} + pg^{\mu \nu}$ where
$\epsilon,p,\rho$ and $v^\mu$ are the mass energy density, pressure, the fluid density 
and the four velocity of the accretion flow, respectively. Throughout this work, 
we will use $G=M_{BH}=c=1$, $M_{BH}$ being the mass of the black hole, other 
symbols represents the known constants. Any representative length and velocity 
will be normalized by $GM_{BH}/c^2$ and $c$, respectively. A polytropic equation 
of state of the form $p=K\rho^\gamma$, $\gamma$ being the ratio of the specific 
heats at constant pressure and at constant volumes, respectively, will be used 
to describe the accreting fluid. The specific enthalpy is defined as 
$h = \dfrac{\epsilon + p}{\rho}$. The polytropic sound speed 
$c_{\rm s} = \sqrt{\left(\dfrac{\partial p}{\partial \epsilon}\right)_{\rm s}}$, 
$s$ being the specific entropy, can thus be obtained as
$c_{\rm s} = \sqrt{\dfrac{\gamma k (\gamma - 1)}{\gamma k + (\gamma -1)(\rho)^{1-\gamma}}}$. 
The normalization condition $v^{\mu}v_{\mu}=-1$ gives 
$v^{t}= \dfrac{\sqrt{v^{2}+f+(v^{\phi})^{2} r^{2}f}}{f}$. 

We consider low angular momentum inviscid accretion in our work. We show that the effective potential for fluid accretion in the 
Schwarzschild metric is $V_{eff} = \sqrt{\dfrac{r^{2}(r-2)}{\lambda^{2}(r-2)+r^{3}}}$, $\lambda$ being 
the specific angular momentum of the flow. If $\lambda < 3.674$ 
matter plunges through the event horizon even if one neglects 
the viscous transport of the angular momentum~\cite{Chakrabarti:1996cc}. Such
a choice of sub Keplarian angular momentum distribution 
prefers that the geometric configuration of matter is of
conical type (see, e.g., \cite{Pratik and Tapas PRD submitted paper} and references therein
for details about various geometrical flow configurations accreting matter can 
have), and any flow variable of the fluid $F$, is assumed to be a slowly varying function of the local flow 
thickness. The vertical averaging of $F$ or any other accretion variable can 
be performed by integrating that variable over $\theta$ and $\phi$ as 
$\int d \theta d \phi \sqrt{-g} F = 4 \pi F  H_{\theta}$, where $H_{\theta}$ is 
the characteristic angular scale of the flow about the equatorial plane and is 
assumed to be the same for all averaged flow variables. The continuity equation 
$(\rho v^{\mu})_{; \mu}=0$ can thus be expressed as
\begin{equation}
\partial_{t} \left( \rho v^{t} \sqrt{-g} H_{\theta} \right) + 
\partial_{r} \left( \rho v \sqrt{-g} H_{\theta} \right) = 0.
\label{t5}
\end{equation}
Finding the off-equatorial accretion solution is beyond the scope of this 
work and we concentrate on the accretion flow on the equatorial plane only.
Accordingly, the continuity equation finally takes the form
\begin{equation}
\partial_{t} \left( \rho v^{t} r^{2} \right) + 
\partial_{r}\left( \rho v r^{2} \right) = 0.
\label{p2a}
\end{equation}
With the help of the continuity equation, the linear energy momentum conservation 
equation on the equatorial plane can be obtained as
\begin{equation}
v^{\nu} \partial_{\nu} v^{\mu} + 
v^{\nu} v^{\lambda} \Gamma^{\mu}_{\nu \lambda} + 
\dfrac{c_{\rm s}^2}{\rho}\left( v^{\mu} v^{\nu} + 
g^{\mu \nu} \right) \partial_{\nu} \rho = 0,
\label{t12}
\end{equation}
where $\Gamma^{\mu}_{\nu \lambda} = \dfrac{1}{2}g^{\mu \eta} \left(\partial_{\nu}g_{\eta \lambda}+\partial_{\lambda}g_{\eta \nu}-\partial_{\eta}g_{\nu \lambda} \right)$. \\
For $\mu=t$, using the vertically averaged accretion variables one can have
\begin{eqnarray}
v^{t} \partial_{t} v^{t} + v \partial_{r} v^{t} + 
\dfrac{v^{t}v}{f} \partial_{r} f + 
\dfrac{c_{\rm s}^{2}}{\rho} \left[ \left((v^{t})^{2}-
\dfrac{1}{f} \right) \partial_{t} \rho + v v^{t} \partial_{r} \rho \right] = 0.
\label{p3}
\end{eqnarray}
For $\mu=r$, the relativistic Euler equation is found to be  
\begin{equation}
v^{t}\partial_{t} v + v \partial_{r} v + 
\dfrac{1}{2} \partial_{r}f \left(1+(v^{\phi})^{2}r^{2} \right) - 
(v^{\phi})^{2} fr+\left( v^{2}+f \right)\dfrac{c_{\rm s}^{2}}{\rho}\partial_{r} \rho 
+ \dfrac{v v^{t}c_{\rm s}^{2}}{\rho} \partial_{t} \rho = 0.
\label{p4}
\end{equation}
Similarly, $\mu = \phi$ provides
\begin{equation}
v^{t} \partial_{t} v^{\phi}+v \partial_{r} v^{\phi} + 
\dfrac{2vv^{\phi}}{r}+ \dfrac{v^{\phi}c_{\rm s}^{2}}{\rho} \left[ v^{t}\partial_{t} \rho 
+ v \partial_{r} \rho \right] = 0.
\label{p5}
\end{equation}

Irrotational  condition can now be introduced by making vorticity operator 
$w_{\mu \nu} = P^{\eta}_{\mu}P^{\eta}_{\nu} \nabla_{\nu} v_{\eta}=0$, where 
$P^{\mu}_{\nu} = \delta^{\mu}_{\nu}-v^{\mu}v_{\nu}$ is the 
projection operator which projects an arbitrary vector in space-time into its 
component in the subspace orthogonal to $v^{\mu}$. Hence for irrotational flow, 
the Euler equation provides 
$\partial_{\mu} \left( h v_{\nu} \right) - 
\partial_{\nu} \left( h v_{\mu} \right) = 0$ and 
for $\left[\left(\mu=r,\nu=t\right),\left(\mu=r,\nu=\phi\right),\left(\mu=t,\nu=\phi\right)\right]$
provides
\begin{eqnarray}
\dfrac{1}{f} \partial_{t} v + \partial_{r} \left( f v{t} \right) 
+\dfrac{c_{\rm s}^{2}}{\rho}\left[\dfrac{v}{f}\partial_{t} \rho + 
v^{t} f \partial_{r} \rho \right] &=& 0,\nonumber \\
f \left[\partial_{r} \left( v^{\phi} r^{2} \right)+ 
v^{\phi} r^{2} \dfrac{c_{\rm s}^{2}}{\rho} \partial_{r} \rho \right] &=& 0,\nonumber \\
\partial_{t} \left[ v^{\phi} g_{\phi \phi} \dfrac{\epsilon +p}{\rho} \right] &=& 0.
\label{p9}
\end{eqnarray}
The last part of the above equation implies the conservation of the angular momentum 
since $v_{\phi} \left( \dfrac{\epsilon+p}{\rho} \right) = L$. The second part of the 
above equation helps to write the Euler equation in the following form
\begin{eqnarray}
v^{t}\partial_{t} v + \partial_{r} \left( \dfrac{v^{2}+f+(v^{\phi})^{2}r^{2}f}{2}\right) + 
\dfrac{c_{\rm s}^{2}}{\rho} \left[ v v^{t} \partial_{t} \rho + 
\left( v^{2}+f+v^{\phi^{2}}r^{2}f \right) \partial_{r} \rho\right] = 0.
\label{p11}
\end{eqnarray}

The stationary solution can thus be obtained by integrating the time independent part of the 
Euler and the continuity equations. We find that the total specific energy ${\cal E}=hv_t$
and the angular momentum $L=hv_\phi$ remain constant along the fluid world line. For steady state,
the mass accretion rate $\dot{M}=2 \pi \rho v r^{2}$ is also found to be a first 
integral of motion. 
\section{Linear perturbation analysis in the Schwarzschild metric}
\label{perturbation}
\noindent
The dynamical velocity, flow density and the mass accretion rates are perturbed 
about their background steady state value. $\Psi_0 = \rho_0 v_0 r^{2}$ is defined as 
the background steady state value of the variable $\Psi$, which actually 
is the accretion rate (apart 
from the geometric constants). We write
\begin{equation}
v = v_{0}+v^{'},~
v^{\phi} = v^{\phi}_{0} + v^{\phi^{'}},~
\rho = \rho_{0}+\rho^{'},~
\Psi^{'} = \left( \rho^{'}v_{0}+\rho_{0}v^{'} \right)r^{2},
\label{p17b}
\end{equation}
where the subscript `0' stands for the stationary values. The derivatives of $\rho^{'}$ and $v^{'}$ can be expressed in terms of derivatives of $\Psi^{'}$ 
as
\begin{equation}
\dfrac{\partial_{t} v^{'}}{v_0} = \dfrac{v_{0}^{t} f^{2}}{\left[ f+(v_{0}^{\phi})^{2}r^{2}f(1-c_{\rm s_{0}}^{2})\right]}  \left[\left(v_{0}^{t}- \dfrac{(v_{0}^{\phi})^{2}r^{2}c_{\rm s_{0}}^{2}f}{v_{0}^{t} f^{2}}\right)\dfrac{\partial_{t} \Psi^{'}}{\Psi_0}+v_{0} \dfrac{\partial_{r} \Psi^{'}}{\Psi_0} \right],
\label{p19a}
\end{equation}
\begin{equation}
\dfrac{\partial_{t} \rho^{'}}{\rho_0} = \dfrac{-v_0}{\left[ f+(v_{0}^{\phi})^{2}r^{2}f(1-c_{\rm s_{0}}^{2})\right]} \left[ v_{0} \dfrac{\partial_{t} \Psi^{'}}{\Psi_0} + v_{0}^{t} f^{2} \dfrac{\partial_{r} \Psi^{'}}{\Psi_0} \right].
\label{p19b}
\end{equation}
The differential equation corresponding to the first order linearly perturbed $\Psi$ thus comes 
out to be
\begin{equation}
\partial_{t}\left( h^{tt} \partial_{t} \Psi^{'}\right) + \partial_{t}\left( h^{tr} \partial_{r} \Psi^{'}\right) + \partial_{r}\left( h^{rt} \partial_{t} \Psi^{'}\right)+ \partial_{r}\left( h^{rr} \partial_{r} \Psi^{'}\right)=0
\label{p21},
\end{equation}
where
$h^{tt} =\dfrac{v_{0}}{v_{0}^{t} \Lambda f} \left[\dfrac{(v_{0}^{t})^{2}f^{2}-c_{\rm s_{0}}^{2}(v_{0}^{2}+
(v_{0}^{\phi})^{2}r^{2}f)}{f^{2}}\right],
h^{tr} = h^{rt} =\dfrac{v_{0}}{v_{0}^{t} \Lambda f}\left[v_{0}v_{0}^{t}(1-c_{\rm s_{0}}^{2})\right],$ and 
$h^{rr} = \dfrac{v_{0}}{v_{0}^{t} \Lambda f}\left[v_{0}^{2} - c_{\rm s_{0}}^{2}(v_{0}^{2}+f)\right]$.

We now construct 
\[
f^{\mu \nu} \equiv
-\dfrac{v_{0}c_{\rm s_0}^2}{v_{0}^{t} \Lambda f} \left[ \begin{array}{ccc}
-\dfrac{1}{f}+(v_0^t)^2 \left(1-\dfrac{1}{c_{\rm s_{0}}^{2}}\right) 
& v_{0}v_{0}^{t}\left(1-\dfrac{1}{c_{\rm s_{0}}^{2}}\right)  \\
\\ v_{0}v_{0}^{t}\left(1-\dfrac{1}{c_{\rm s_{0}}^{2}}\right) & f+v_0^2\left(1-\dfrac{1}{c_{\rm s_{0}}^{2}}\right) \\
\end{array} \right],\] where $\Lambda = (1+(v_{0}^{\phi})^{2}r^{2}(1-c_{\rm s_{0}}^{2}))$.
The wave equation in $\Psi^{'}$ can now be written in a more compact invariant way as
\begin{equation}
\partial_{\mu} \left( f^{\mu \nu} \partial_{\nu} \Psi^{'} \right) = 0.
\end{equation}
One may try to identify this with the Klein-Gordon equation so as to define the acoustic metric, but there is an ambiguity associated with divergent conformal factor for defining acoustic metric in $(1+1)$ dimension~\cite{Barcelo:2004wz,bar05}. So we define an effective metric, $f_{\mu \nu}$, which still gives the desired causal structure.
\[
f_{\mu \nu} \equiv
-\dfrac{fv_{0}^t}{v_{0}} \left[ \begin{array}{ccc}
f+v_0^2\left(1-\dfrac{1}{c_{\rm s_{0}}^{2}}\right)& -v_{0}v_{0}^{t}\left(1-\dfrac{1}{c_{\rm s_{0}}^{2}}\right)  \\
\\ -v_{0}v_{0}^{t}\left(1-\dfrac{1}{c_{\rm s_{0}}^{2}}\right) & -\dfrac{1}{f}+(v_0^t)^2 \left(1-\dfrac{1}{c_{\rm s_{0}}^{2}}\right) \\
\end{array} \right],\]
Setting $f_{tt}=0$, gives the sonic point condition $u_0^2=c_{\rm s_0}^2$, where $u$ is the radial velocity in local rest frame.
We thus demonstrate that for the 
background flow of relativistic fluid in strong gravity, 
the analogue spacetime forms an intrinsic manifold configuration of 
which remains invariant for various perturbation schemes.
\section*{Stability analysis in the Schwarzschild metric}
We now study the stability of the stationary solutions using the trial acoustic wave solution $\Psi^{'} = p(r)\exp(-i \omega t)$. Substituting this in the wave equation results in 
\begin{equation}
\omega^{2} p(r) f^{tt} + i \omega [\partial_{r}(p(r)f^{rt})+ f^{tr} \partial_{r} p(r)]-[\partial_{r} (f^{rr} \partial_{r} p(r))] = 0. 
\label{eqn4.52}
\end{equation}
We use the trial power solution 
\begin{equation}
p(r) = \exp \left[ \sum_{k=1}^n \dfrac{k_{n}(r)}{\omega^{n}} \right].
\label{eqn4.30}
\end{equation}
Substitute this in Eqn.(\ref{eqn4.52}) result in a series in $\omega$ and setting the coefficients of individual powers of $\omega$ to zero gives the leading coefficients. Equating the coefficient of $\omega^{2}$ term to zero gives
\begin{equation}
k_{-1} = i \int \dfrac{v_{0}v_{0}^{t}(1-c_{\rm s_{0}}^{2}) \pm c_{\rm s_{0}}\sqrt{\Lambda}}{v_{0}^{2}-c_{\rm s_{0}}^{2}(v_{0}^{2}+f)} dr.
\label{eqn4.31}
\end{equation}
Similarly equating the coefficient of $\omega$ term to zero gives
\begin{equation}
k_{0} = \dfrac{-1}{2} \ln \left(\dfrac{c_{\rm s_{0}} v_{0}}{v_{0}^{t}f\sqrt{\Lambda}} \right).
\label{eqn4.32}
\end{equation}
In the asymptotic limit, we get $k_{-1} \sim r, k_{0} \sim \ln r$ and $k_{1} \sim r^{-1}$. Since $\omega \gg 1$ we can see that $\omega r \gg \ln r \gg \dfrac{\omega}{r}$, the power series converges. Hence the stationary solutions obtained are self-consistent. 
\section{Acoustic geometry for the Rindler spacetime}
\label{rindler}
A uniformly accelerating observer in the Minkowski space is described by the 
Rindler co ordinates \cite{General Relativity}. We now introduce uniform rotation $\phi \rightarrow \phi -\Omega t$ and the line element comes out to be 
$ds^{2} = -(a^{2}x^{2}-\Omega^{2}\rho^{2}) dt^{2}-2 \Omega \rho^{2}dt d \phi + 
\rho^{2} d \phi^{2} + dx^{2} + d \rho^{2}$, where a, and $\Omega$ are constants.
We diagonalise the Rindler metric to find the acoustic geometry through the procedure
what has been followed for the Schwarzschild spacetime. We define 
a Killing vector $\chi_{a}$, which is a linear combination of both 
$(\partial_{t})_{a}$ and $(\partial_{\phi})_{a}$ and is orthogonal 
to both $\partial_{t}$ and $\partial_{\phi}$. Hence
$\chi_{a} = (\partial_{t})_{a} + \alpha (\partial_{\phi})_{a}$. The orthogonality 
property provides $\alpha= \Omega$ resulting 
$\chi_{a} = (\partial_{t})_{a} + \Omega (\partial_{\phi})_{a}$. Hence 
in the new diagonalized basis
$ \left[\chi_{a}, \left( \partial_{x}\right)_{a}, 
\left( \partial_{\phi} \right)_{a}, 
\left( \partial_{\rho} \right)_{a} \right] $, one obtains
$g_{\tau \tau} = -a^{2}x^{2}, g_{xx} = 1, g_{\rho \rho} = 1, g_{\phi \phi} = \rho^{2}$.
The velocity component along the $\chi$, $v^{\tau}$ can be obtained from the normalization condition 
as $v^{\tau} = \dfrac{\sqrt{1+v^{2}+ (\rho v^{\phi})^{2}}}{ax}$. The metric in the transformed 
basis is a static one and hence following the linear perturbation analysis as has been 
done for the Schwarzschild spacetime, one obtains the elements of the effective metric 
for the Rindler spacetime as 
 \[
f_{\mu \nu} \equiv
-\dfrac{ax v_0^{\tau}}{v_0} \left[ \begin{array}{ccc}
1+v_0^2\left(1-\dfrac{1}{c_{\rm s_{0}}^{2}}\right)& -v_{0}v_{0}^{\tau}\left(1-\dfrac{1}{c_{\rm s_{0}}^{2}}\right)  \\
\\ -v_{0}v_{0}^{\tau}\left(1-\dfrac{1}{c_{\rm s_{0}}^{2}}\right) & -\dfrac{1}{ax}+(v_0^{\tau})^2 \left(1-\dfrac{1}{c_{\rm s_{0}}^{2}}\right) \\
 \end{array} \right].\]

\section*{Acknowledgements}
\noindent
Long term visit of DBA at HRI has been supported by the planned project fund of the 
Cosmology and the High Energy Astrophysics subproject of HRI.

\end{document}